\documentclass[a4paper,11pt]{article}
\usepackage{jinstpub} 
\usepackage{graphicx}
\usepackage{subcaption}
\usepackage{tabularx}
\usepackage{array} 

\title{\boldmath Shielding design and construction for the XINGZHI cold neutron triple-axis spectrometer: integrating velocity selector, S-bender polarizer, and neutron guide}
\author[a]{Hongxia Zhang,}
\author[a]{Daye Xu,}
\author[b]{Florian Gr\"{u}nauer,}
\author[a]{Jinchen Wang,}
\author[c]{Yuntao Liu,}
\author[a,1]{Juanjuan Liu,\note{Corresponding author.}}
\author[a,2]{Peng Cheng,\note{Corresponding author.}}
\author[d,e,3]{Wei Bao\note{Corresponding author.}}

\affiliation[a]{Laboratory for Neutron Scattering, School of Physics, Renmin University of China, Beijing 100872, China}
\affiliation[b]{Physics Consulting, Zorneding 85604, Germany}
\affiliation[c]{China Institute of Atomic Energy, PO Box 275-30, Beijing 102413, China}
\affiliation[d]{Department of Physics, City University of Hong Kong, Kowloon, Hong Kong SAR}
\affiliation[e]{Center for Neutron Scattering, City University of Hong Kong, Kowloon, Hong Kong SAR}
\emailAdd{juanjuanliu@ruc.edu.cn, pcheng@ruc.edu.cn, weibao@cityu.edu.hk}

\abstract{The optimized design and construction of shielding for the velocity selector, S-bender polarizer, and neutron guide in the XINGZHI cold neutron triple-axis spectrometer have been successfully completed. The challenge of limited space available for XINGZHI has been met by a compact shielding structure, developed through Monte Carlo simulations and realized by purpose-built mechanical design, thus leaving more floor space for the spectrometer. An innovative lead-antimony alloy ensures rigidity for self-support, while optimized tolerances for gaps between mating blocks meet safety, machinability, and mounting requirements. Additionally, a long-distance alignment method has been implemented, facilitating the alternating installation of the neutron guide and shielding system. After six years of operation, this structure has been verified to meet radiation safety standards while maximizing available space, marking a significant advancement in neutron instrumentation in China and offering valuable methodologies for future development.}


\begin{document}

\maketitle
\section{Introduction}
The construction of the cold neutron triple-axis spectrometer (CNTAS), now named XINGZHI, commenced in 2013 at the China Advanced Research Reactor (CARR) \cite{CHENG201617}. Having been in successful operation for six years, XINGZHI is the first instrument of its kind in China capable of inelastic neutron scattering studies with polarization analysis.
The CARR is designed to work at 60~MW, with its compact core providing an unperturbed maximum thermal neutron flux of 8~$\times$~10$^{14}$~n~cm$^{-2}$~s$^{-1}$, and features a cold source installed inside the reflector \cite{CHEN2006966}. 
XINGZHI is installed at the end of the cold neutron guide A (CNGA). The CNGA has a cross section of 30~mm (width)~$\times$~150~mm (height), and consists of a 17\mbox{-m}-long curved guide with a cut-off wavelength of 0.2~nm and a straight guide, with an $m$-value of 2. The multiplexing cold neutron spectrometer BOYA \cite{Wang_2025} is positioned at the first beam port, upstream of XINGZHI. The primary optical assembly of XINGZHI begins with a velocity selector inserted into the beam line to remove higher-order neutrons. This is followed by an optional S-bender polarizer, which can be interchanged with a normal section of straight guide for unpolarized experiments. Finally, a 3.3\mbox{-m}-long neutron guide transmits the beam to the XINGZHI monochromator \cite{CHENG201617}.

The design of XINGZHI required comprehensive simulations addressing not only neutron optics for optimizing flux and momentum and energy resolution \cite{CHENG201617,Wang_2025,Zhang_2024} but also biological shielding.
Effective shielding is essential to minimize detector background and to ensure safety. Also, compact shielding is essential to maximize the available space for advanced instrumentation, and accommodate as many instruments as possible along the beam line. Furthermore, floor load constraints must be carefully considered in the design of the biological shielding, specifically the 7~tons/m$^2$ constraint at the CARR guide hall.

To meet radiological and mechanical constraints, various shielding materials and structures have been explored in the field, including the combination of different density concretes \cite{Pyka2002,YILMAZ20112204}, the mixture of steel resin, polyethylene, and a boron compound \cite{CALZADA201177}, the combination of lead and polyethylene or steel \cite{Grunauer_2023,GRUNAUER2023167854,ERSEZ20061268,OSBORN20061321}. Building upon these material studies to achieve optimal compactness, a specialized sandwich shielding structure was developed and successfully implemented for the monochromator shielding of XINGZHI and BOYA \cite{Wang_2024}.

However, radiation protection extends beyond the monochromator. The upstream components, the velocity selector, the S-bender polarizer, and the connecting neutron guides, present radiological and engineering challenges in some aspects even more complex than those of the monochromator.  
The presence of Gadolinium (Gd) in the S-bender polarizer made by Mirrotron Ltd., Hungary creates an intense source of high-energy capture gammas, which in turn generates a secondary fast neutron background via the ($\gamma$, n) photo-nuclear effect in lead shielding.

To address this compounded radiation problem, a tailored B$_{4}$C\slash lead\slash borated polyethylene (BPE) sandwich shielding structure was invented. Furthermore, to overcome the poor structural integrity of pure lead shielding for the long neutron guide, lead-antimony alloy is innovatively employed. Combined with modular designs featuring labyrinth-type gaps, the system ensures radiation tightness while maintaining accessibility for maintenance. A robust methodology for installing and aligning the long beam line components and shielding is presented. The completed shielding system was successfully installed and validated at CARR. The radiation protection rules specify that an annual personal dose of 3~mSv may not be exceeded. This rule can be followed by reduction of the local dose rate by appropriate radiation shielding and minimizing the duration of stay in regions with increased radiation levels. An internal guideline of the facility specifies that for distances bigger than 50~cm from the outer surfaces of a radiation shielding, a total dose rate of 6~$\upmu$Sv/h (3~$\upmu$Sv/h for neutron radiation and 3~$\upmu$Sv/h for gamma radiation) should not be exceeded in freely accessible areas, and if this is not possible, the access to the area with increased radiation level has to be limited. To ensure personnel safety during normal operation, no non-shielded beam line components are located in these accessible areas. Measurements confirmed that the dose rates outside the shielding were well below the design limit during the six years of operation.

Furthermore, the efficacy and advanced nature of both monochromator and neutron guide shielding with this compact design have been further demonstrated by its subsequent adoption for the HETU diffractometer at the China Mianyang Research Reactor (CMRR) \cite{Wang:te5115}. Motivated by its proven long-term reliability and broader applicability, we present the detailed engineering design here to provide a validated reference for the future construction of advanced neutron instruments.
The subsequent sections of this paper are organized as follows: Section 2 details the Monte Carlo optimization,  Section 3 presents the mechanical designs and material choices, and Section 4 describes the installation procedures and radiation monitoring measurements.

\section{Monte Carlo simulations}
In this section the results of the Monte Carlo simulations are described. The Monte Carlo code MCNP5 \cite{mcnp5_manual} was applied. All results were obtained by track-length-estimators. As variance reduction methods implicit capture, geometry splitting, and adjustment of weight cutoffs were applied. All dose rates were obtained by converting the energy dependent fluxes using the ICRP74 fluence-to-dose conversion factors $h^*(10)$ \cite{ICRP74-1}.

\subsection{Virtual neutron sources}

\begin{figure}[!htbp]
	\centering
	\includegraphics[width=1\columnwidth]{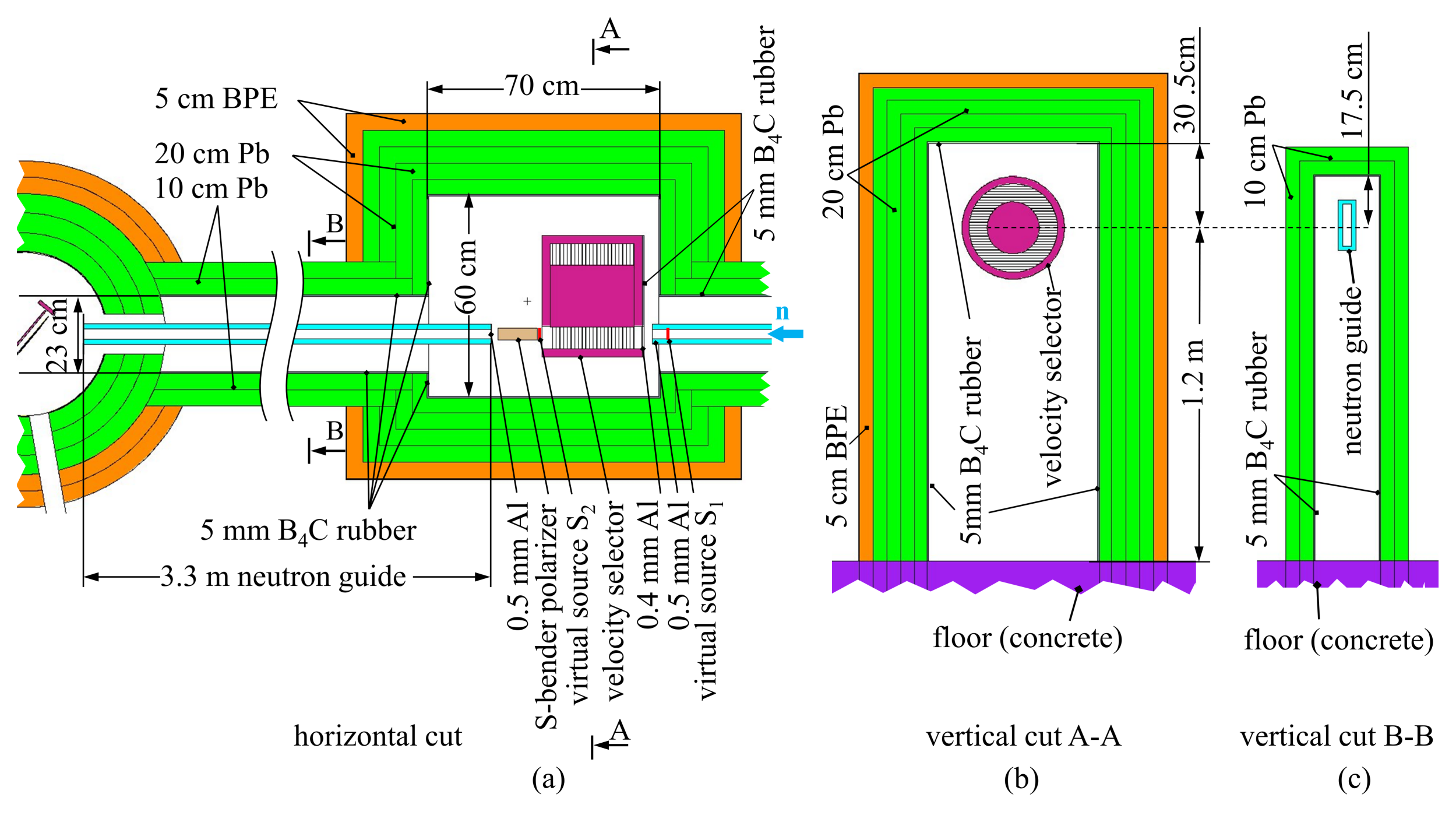}
	\caption{Location of the virtual neutron sources S$_{1}$ and S$_{2}$. (a) Horizontal cut along the neutron beam axis, and vertical cut perpendicular to the neutron beam axis through the Monte Carlo model of (b) the velocity selector shielding and (c) the neutron guide shielding. The incoming neutron beam is depicted by blue arrow.}
	\label{fig1}
\end{figure}

To represent the beam characteristics, two virtual neutron area sources are used in the MCNP simulations. These source terms were derived from our previous ray-tracing simulations for the XINGZHI spectrometer \cite{CHENG201617} using the McStas code \cite{McStas1999}.
The first virtual neutron source S$_{1}$ is located in front of the velocity selector, and the second virtual neutron source S$_{2}$ that represents the transmission of the desired neutron energy band through the velocity selector, is located behind the absorber blades in front of the aluminum-exit-window of the velocity selector, see figure~\ref{fig1}. 
Both sources cover the full inner cross section of the neutron guide (30~$\times$~150~mm$^{2}$). 
For S$_{1}$, the total source yield is assumed to be 3.2~$\times$~10$^{11} $~n~s$^{-1}$ resulting in a neutron flux of 7.1~$\times$~10$^{9} $~n~cm$^{-2}$~s$^{-1}$. For S$_{2}$, its source yield is 2.96~$\times$~10$^{10} $~n~s$^{-1}$ resulting in a neutron flux of 6.58~$\times$~10$^{8} $~n~cm$^{-2}$~s$^{-1}$. These source yields were derived assuming the CARR operates at its full power of 60~MW. Since the routine operational power of CARR is 30~MW, this assumption provides a safety factor of two for the subsequent dose rate evaluations.

The spectral neutron flux intensity from McStas is transferred to MCNP5. Neutrons are emitted in an angle interval between 0$^{\circ}$ and 0.9$^{\circ}$ around the beam axis. This $0.9^\circ$ value was derived as the effective average of the divergence limits calculated by McStas ($0.8^\circ$ horizontal and $1.0^\circ$ vertical). Neutrons are emitted isotropically within this angular interval around the beam axis.
Since the source S$_{1}$ is applied just in front of the Al window and the velocity selector, neutrons would cross the same structures as long as the angle remains within the geometric acceptance angle. Hence the simulation is not sensitive to this angle. The same holds for the source S$_{2}$ in front of the S-bender polarizer. For S$_{1}$, the transmission of the desired energy band of neutrons through the velocity selector is not considered (in reality not the whole neutron beam is absorbed: roughly 10\% is transmitted). This conservative assumption maximizes the local production of radiation, defining the upper limit for the shielding requirements around this component.

The velocity selector made by Astrium GmbH, Germany is a standard selector with the rotor outer diameter of 290~mm and the rotor length of 250~mm, and blades coated with boron. The front surface of the velocity selector is covered by a boron carbide rubber mat in the simulations, except the entrance window for the neutrons. In the simulations, boron carbide rubber mat was modeled as B$_{4}$C with a density of 0.9~g~cm$^{-3}$.
The S-bender polarizer (Mirrotron Ltd., Hungary) consists of a stack of thin silicon wafers, coated on both sides with Fe-Si supermirrors and an absorbing Gd layer. The desired spin-component passes the polarizer and is not absorbed in the polarizer. However, in the simulations the different structures of the S-bender polarizer are assumed to be homogenized with 50.8~wt.\% Si, 45.7~wt.\% Fe, and 3.4~wt.\% Gd. 
In the homogenized model, the Gd atoms are distributed uniformly throughout the entire volume. The mass density of the polarizer was not adjusted to reproduce the real neutron transmission. Consequently, the probability of neutron capture and subsequent gamma production is increased in the simulations compared to the reality where neutrons can bypass the absorbers. This represents the worst case of a misaligned S-bender polarizer. 

\subsection{Radiation shielding for the velocity selector and S-bender polarizer}

\begin{figure}[!htbp]
	\centering
	\begin{subfigure}[b]{1\textwidth}
		\centering
		\includegraphics[width=\textwidth]{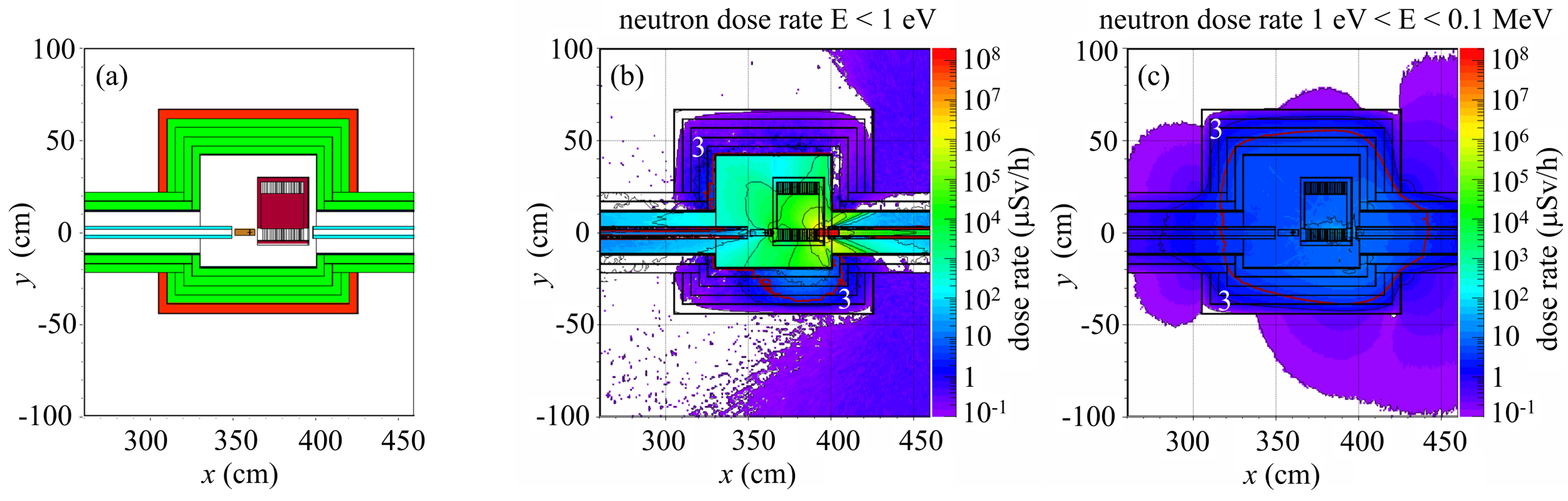}
	\end{subfigure}
	\vspace{2em}
	\begin{subfigure}[b]{1\textwidth}
		\centering
		\includegraphics[width=\textwidth]{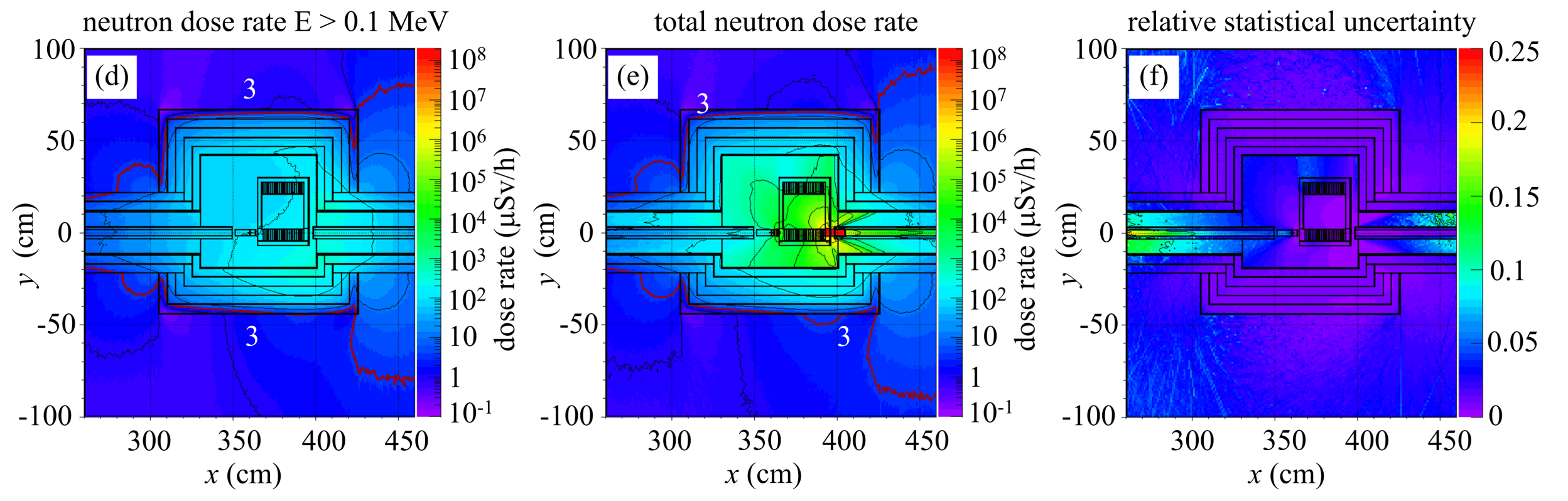}
	\end{subfigure}
	\caption{Distribution of the neutron dose rate. (a) Horizontal area through the velocity selector shielding along the beam axis for which the radiation distributions are shown in the following images, (b) neutron dose rate in the energy region E~$<$1~eV, (c) neutron dose rate in the energy region 1~eV$<$~E~$<$0.1~MeV, (d) neutron dose rate in the energy region E~$>$~0.1~MeV, (e) total neutron dose rate, and (f) relative statistical uncertainty of the total neutron dose rate. The dark-red line is the 3~$\upmu$Sv/h border. Values below 0.1~$\upmu$Sv/h are plotted in white. The spatial resolution is $\Delta x = 1~\mathrm{cm}$, $\Delta y = 1~\mathrm{cm}$, and $\Delta z = 10~\mathrm{cm}$.}
	\label{fig2a}
\end{figure}

\begin{figure}[!htbp]
	\centering
	\begin{subfigure}[b]{1\textwidth}
		\centering
		\includegraphics[width=\textwidth]{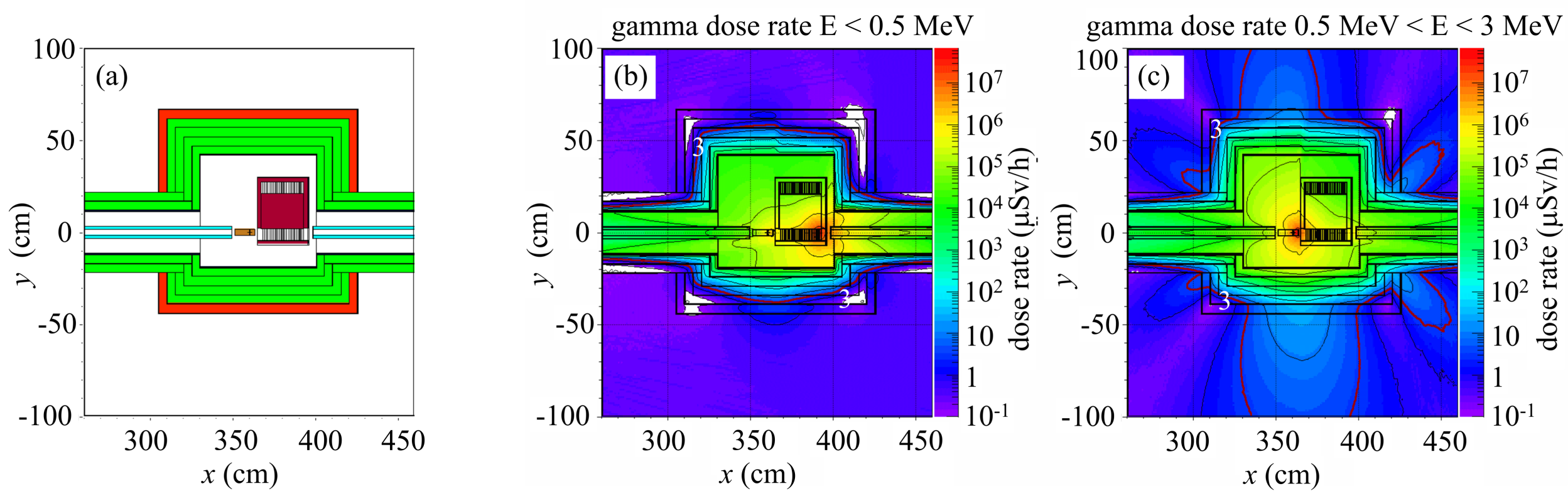}
	\end{subfigure}
	\vspace{2em}
	\begin{subfigure}[b]{1\textwidth}
		\centering
		\includegraphics[width=\textwidth]{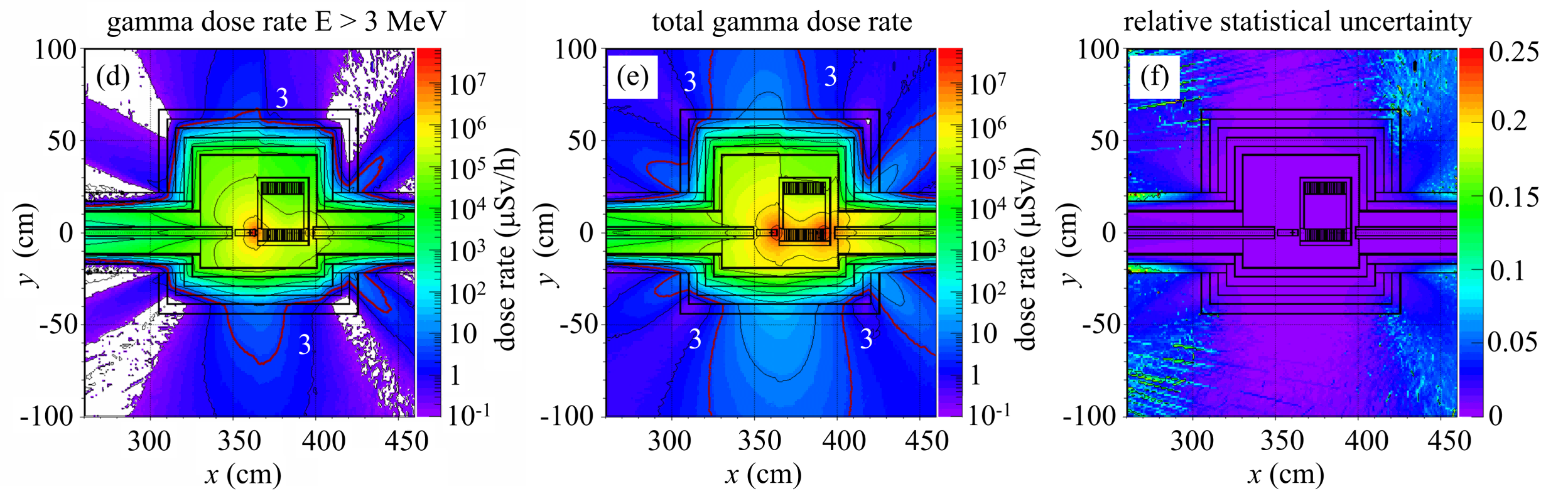}
	\end{subfigure}
	\caption{Distribution of the gamma dose rate. (a) Horizontal area through the velocity selector shielding along the beam axis for which the radiation distributions are shown in the following images, (b) gamma dose rate in the energy region E~$<$~0.5~MeV, (c) gamma dose rate in the energy region 0.5~MeV~$<$~E~$<$~3~MeV, (d) gamma dose rate in the energy region E~$>$~3~MeV, (e) total gamma dose rate, and (f) relative statistical uncertainty of the total gamma dose rate. The dark-red line is the 3~$\upmu$Sv/h border. Values below 0.1~$\upmu$Sv/h are plotted in white. The spatial resolution is $\Delta x = 1~\mathrm{cm}$, $\Delta y = 1~\mathrm{cm}$, and $\Delta z = 10~\mathrm{cm}$.}
	\label{fig3a}
\end{figure}
Several shielding layouts with different thicknesses and materials were tested by simulations for optimization of the shielding. In the following only the results for the final layout are described. As illustrated in figure~\ref{fig1} (b), the shielding consists of (from inside to outside): 5\mbox{-mm}-thick boron carbide rubber mat, 200\mbox{-mm}-thick lead and 50\mbox{-mm}-thick borated polyethylene layer (BPE) with 5~wt.\% B$_{2}$O$_{3}$. The simulation results for neutron radiation and gamma radiation are presented in figure~\ref{fig2a} and figure~\ref{fig3a} in a horizontal plane at the beam axis level, respectively. Virtual sources S$_{1}$ and S$_{2}$ are applied. Most thermal neutrons from virtual source S$_{1}$ are absorbed in the boron coating of the absorber blades in the velocity selector (figure~\ref{fig2a} (b)). They are captured by the reaction $^{10}\text{B}(\mathrm{n}, \alpha)^{7}\text{Li}$ with emission of gamma radiation with an energy of 0.478~MeV \cite{IAEA_PromptGamma_2007}.
Therefore the velocity selector becomes an intense source of low-energy gamma radiation (figure~\ref{fig3a} (b)). For gamma radiation in this energy range, the attenuation of the surrounding lead shielding is quite high and the thresholds for photonuclear reactions are not exceeded.

However, a small amount of thermal neutrons is scattered or absorbed in the Al exit window of the incoming neutron guide (0.5~mm thickness) or the Al entrance window of the velocity selector (0.4~mm thickness). Most of the scattered thermal neutrons are absorbed in the inner boron carbide rubber mat of the shielding also by the reaction $^{10}\text{B}(\mathrm{n},\alpha)^{7}\text{Li}$ (figure~\ref{fig2a} (b)). Only low-energy gamma radiation is produced and the thermal neutron flux in the adjacent lead layer is minimized which keeps the activation of the lead layer low \cite{Grunauer_LaDiff_2024}. Neutrons that are absorbed in the Al windows produce high-energy gamma radiation up to 7.7~MeV \cite{IAEA_PromptGamma_2007} by the reaction $^{27}\text{Al}(\mathrm{n},\gamma)^{28}\text{Al}$. Production of this high-energy gamma radiation cannot be prevented. The same holds for the S-bender polarizer where gamma radiation with an energy of up to 6.75~MeV \cite{IAEA_PromptGamma_2007} is produced by the neutron capture reaction $^{157}\text{Gd}(\mathrm{n},\gamma)^{158}\text{Gd}$ (figure~\ref{fig3a} (d)). This gamma radiation exceeds the threshold for the photonuclear reaction $(\gamma, \mathrm{n})$ in lead (6.7~MeV for $^{207}\text{Pb}$ and 7.4~MeV for $^{208}\text{Pb}$ \cite{IAEA_Photonuclear_2000}). Hence the lead shielding becomes a source of fast neutrons (figure~\ref{fig2a} (d)). The simulations showed that the 50\mbox{-mm}-thick BPE layer on the outside of the shielding is necessary to attenuate the fast neutron dose rate to the desired level.

\begin{figure}[!htbp]
	\centering
	\begin{subfigure}[b]{1\textwidth}
		\centering
		\includegraphics[width=\textwidth]{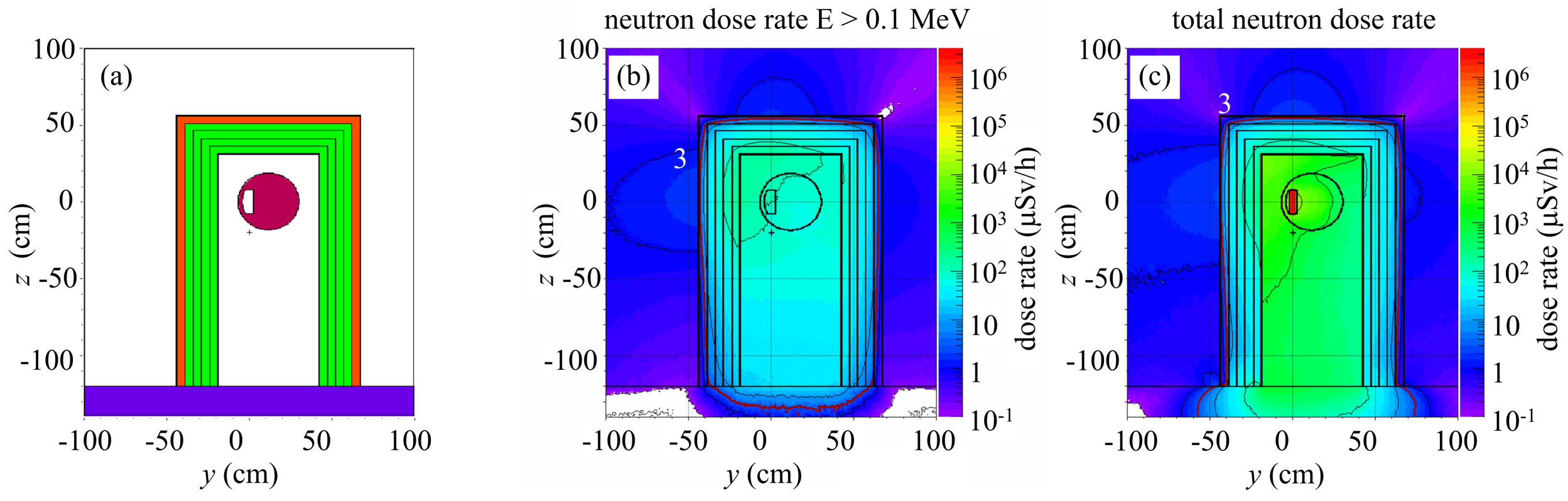}
	\end{subfigure}
	\vspace{2em}
	\begin{subfigure}[b]{1\textwidth}
		\centering
		\includegraphics[width=\textwidth]{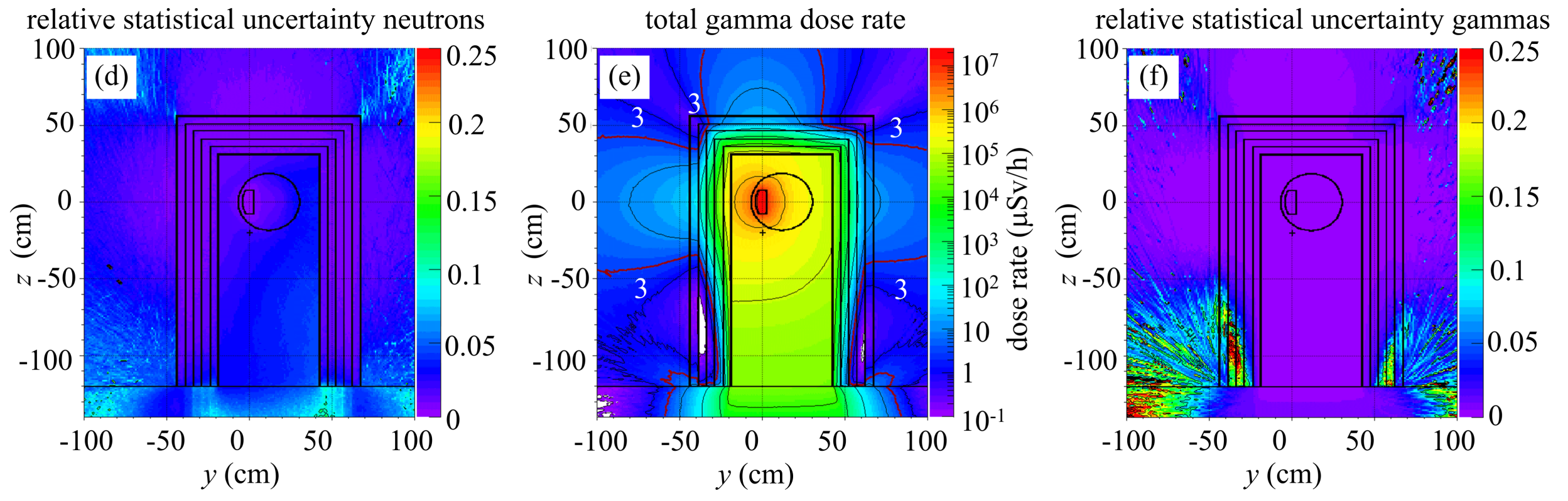}
	\end{subfigure}
	
	\caption{Distribution of the neutron and gamma dose rates in a vertical area. (a) Vertical area through the velocity selector shielding perpendicular to the beam axis (at $x = 360$~cm) for which the radiation distributions are shown in the following images, (b) neutron dose rate in the energy region E~$>$~0.1~MeV, (c) total neutron dose rate, (d) relative statistical uncertainty of the total neutron dose rate, (e) total gamma dose rate, and (f) relative statistical uncertainty of the total gamma dose rate. The dark-red line is the 3~$\upmu$Sv/h border. Values below 0.1~$\upmu$Sv/h are plotted in white. The spatial resolution is $\Delta x = 10~\mathrm{cm}$, $\Delta y = 1~\mathrm{cm}$, and $\Delta z = 1~\mathrm{cm}$.}
	\label{fig4a}
\end{figure}

The simulation results for a vertical plane through the shielding of the velocity selector perpendicular to the beam axis (at $x = 360$~cm) are shown in figure~\ref{fig4a}.
A minor scattering of gamma radiation through the floor (concrete) is observable, but its localized dose rate remains well below $1$~$\upmu$Sv/h, making its contribution to the overall radiation field negligible.
Furthermore, the simulated neutron dose rate remains below the required limit in freely accessible areas (figure~\ref{fig4a} (c)).

As observed from the horizontal and vertical profiles (figures~\ref{fig3a} (e) and \ref{fig4a} (e)), localized gamma dose rates at the 50~cm boundary of up to 6~$\upmu$Sv/h are obtained. The neutron dose rate keeps below $3$~$\upmu$Sv/h at the 50~cm boundary around the velocity selector shielding except around the incoming neutron guide shielding where up to $4$~$\upmu$Sv/h are obtained at the 50~cm boundary (figure~\ref{fig2a} (e)).

Nevertheless, it was decided to realize this shielding layout based on the following comprehensive evaluations:

(1) The simulated case is a worst-case scenario for a reactor power of 60~MW. 
Thus, under the CARR’s routine 30~MW operation, the localized gamma dose rate scales down to 3~$\upmu$Sv/h.

(2) In the simulations the worst-case scenario of a misaligned polarizer is assumed.

(3) In the simulations, the shielding layout for the incoming neutron guide was assumed to be the same as that of the outgoing guide (figure \ref{fig1}). The incoming neutron guide shielding was not subject of the simulations presented here. The neutron flux and the neutron loss in the incoming neutron guide (white beam) are bigger compared to the outgoing guide. The incoming guide shielding is therefore in reality thicker and contains a borated polyethylene layer for reduction of neutron radiation. Hence the simulations overestimate the dose rate around the incoming guide shielding.  Even with this overestimation, the neutron dose rate drops to 2~$\upmu$Sv/h under routine 30~MW operation. 

(4) The annual dose limit is satisfied not only by physical shielding but also by limiting the duration of stay and applying strict administrative controls. In the real installation, one side of the shielding ($+ y$ direction) is located in a narrow space adjacent to another beam line, and is not a normal personnel-access route. On the spectrometer side ($- y$ direction), physical barriers and warning signs are used to control access to the localized regions identified by the conservative 60~MW calculation. Furthermore, the XINGZHI spectrometer is operated remotely, and necessary personnel access to the spectrometer area is limited to short intervals, under the applicable radiation-protection procedures.

(5) If routine 60~MW operation with the polarizer translated into the beam is implemented in the future, additional local lead shielding will be installed in
the reserved space as required, and a confirmatory radiation survey will be performed before routine access to the affected regions is permitted.

(6) The instrument will not be operated with the polarizer all the time but also without the polarizer. As shown below lower dose rates are obtained in this case.
 
\begin{figure}[!htbp]
	\centering
	\includegraphics[width=0.7\columnwidth]{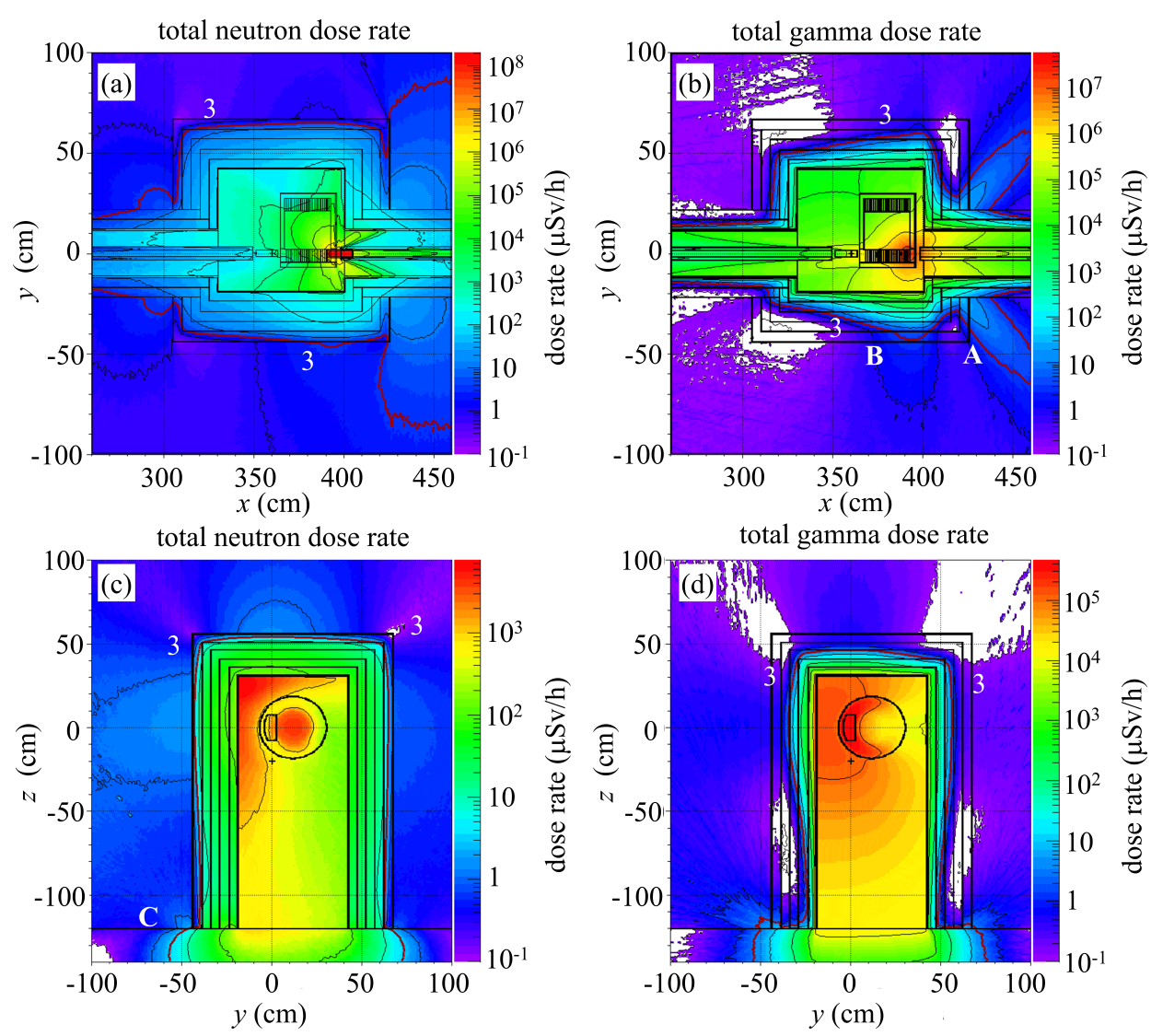}
	\caption{Neutron and gamma dose rate distributions with only the virtual source S$_{1}$ applied. (a)~total neutron dose rate and (b)~total gamma dose rate in the horizontal plane along the beam axis. (c)~total neutron dose rate and (d)~total gamma dose rate in the vertical plane perpendicular to the beam axis (at $x = 360$~cm). The dark-red contour indicates the 3~$\upmu$Sv/h border. Values below 0.1~$\upmu$Sv/h are plotted in white. All relative statistical uncertainties are below 10\%. The spatial resolution is $\Delta x = 1~\mathrm{cm}$, $\Delta y = 1~\mathrm{cm}$, and $\Delta z = 10~\mathrm{cm}$ for (a) and (b). The spatial resolution is $\Delta x = 10~\mathrm{cm}$, $\Delta y = 1~\mathrm{cm}$, and $\Delta z = 1~\mathrm{cm}$ for (c) and (d).}
	\label{fig5b}
\end{figure}

In the simulations, this polarizer-free operation is represented by applying only virtual source S$_{1}$. The simulation results for this case are presented in figure~\ref{fig5b}.
In this configuration, the localized gamma dose rate peaks beyond the 50~cm boundary are eliminated, dropping below the 3~$\upmu$Sv/h limit (figure~\ref{fig5b} (b) and (d)), as neutron capture by Gd is avoided.
The 3~$\upmu$Sv/h border line retracts inward by 5~cm (figure~\ref{fig5b}~(a)), compared to the full configuration. This minor spatial shift confirms that the gammas emitted by Gd have a negligible influence on the photonuclear reactions in the lead shielding. Consequently, most photonuclear reactions are caused by the high-energy gamma radiation from neutron capture in the Al window at the end of the incoming neutron guide and in the entrance Al window of the velocity selector.
 
\subsection{Radiation shielding for the neutron guide}
\begin{figure}[!htbp]
	\centering
	\includegraphics[width=0.5\columnwidth]{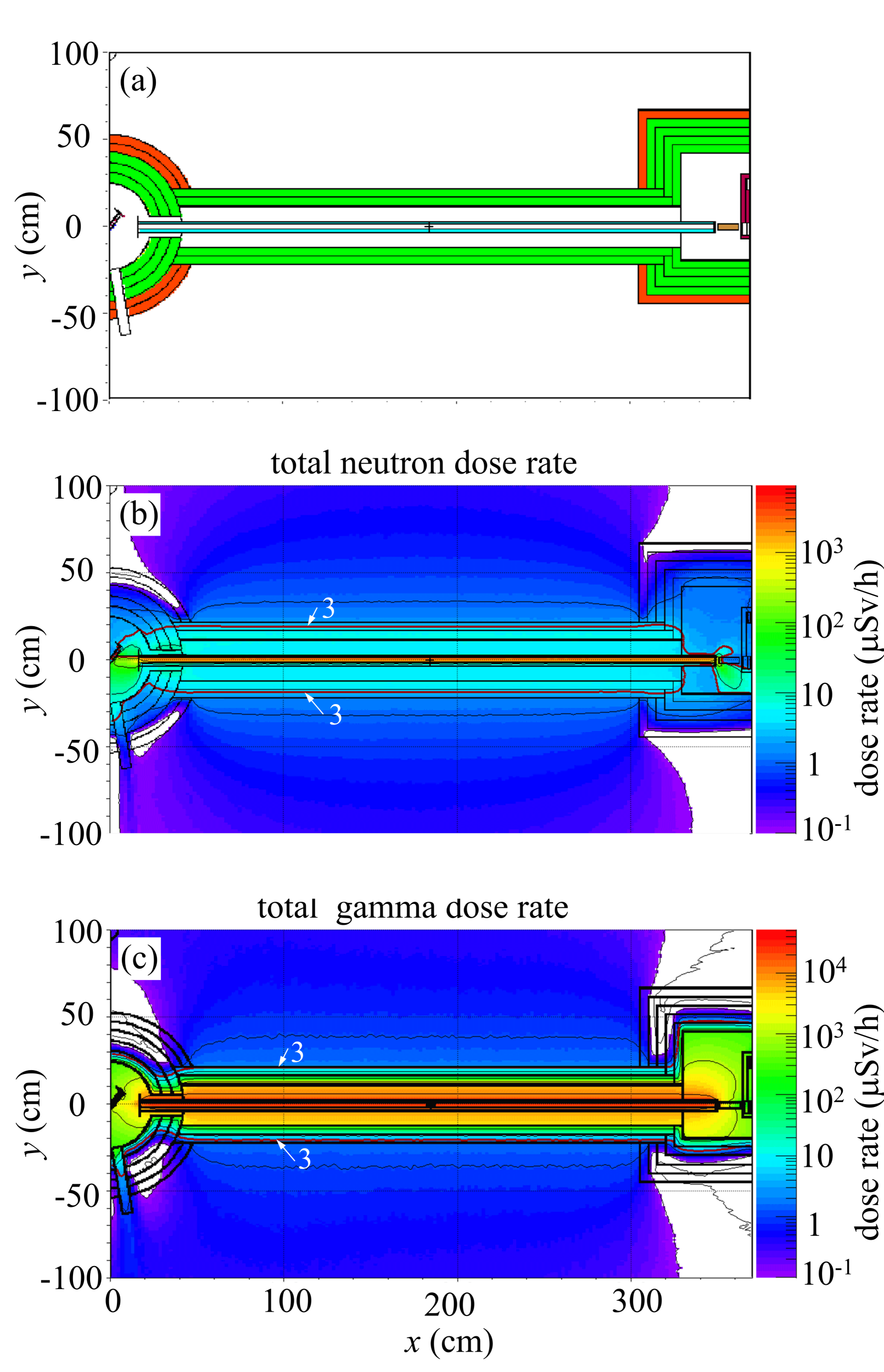}
	\caption{Neutron and gamma dose rate distributions. (a)~Horizontal area through the neutron guide shielding along the beam axis for which the radiation distributions are shown in the following images, (b)~total neutron dose rate, and (c)~total gamma dose rate. Values below 0.1~$\upmu$Sv/h are plotted in white. All relative statistical uncertainties are below 10\%. The spatial resolution is $\Delta x = 2~\mathrm{cm}$, $\Delta y = 1~\mathrm{cm}$, and $\Delta z = 10~\mathrm{cm}$. In this simulation the lost neutrons, that are not total-reflected by the guide, the generated gamma radiation and neutrons from photonuclear reactions are considered.}
	\label{fig5a}
\end{figure}

Downstream of the velocity selector and S-bender, a 3.3\mbox{-m}-long neutron guide made by Mirrotron Ltd., Hungary transmits the neutron beam to the XINGZHI monochromator.
The neutron guide walls consist of Borofloat33 glass (Schott Jena, Germany). The boron content reduces the production of high-energy gamma radiation by neutron absorption. In the simulations, a thickness of the side guide walls of 16~mm and a thickness of 14.8~mm for the top and bottom walls are assumed. The inner coating of the guide consists of NiMo (0.6~$\upmu$m) and Ti (0.6~$\upmu$m) with an $m$-value of 2.

A small amount of the neutrons in the neutron guide is not reflected by the inner surfaces of the guide but absorbed or scattered in the inner coating (NiMo/Ti) of the guide or the guide walls (Borofloat33 glass). Due to the gamma production by neutron absorption a shielding around the guide is necessary. In the simulations, a neutron source along the inner surfaces of the 3.3\mbox{-m}-long guide is used. It should be mentioned that the source yield is 8.59~$\times$~10$^{9}$~n/s corresponding to the total loss of neutrons in the guide, which is calculated through McStas simulation \cite{CHENG201617}. Neutrons that are total-reflected in the guide are of no interest for the shielding because they remain inside the guide and produce no gamma radiation.

In the simulations, a shielding made of lead around the neutron guide covered by a 5\mbox{-mm}-thick boron carbide rubber mat with a B$_{4}$C concentration of 0.9~g~cm$^{-3}$ is considered. The simulations showed that a lead shielding of 100~mm thickness reduces the gamma dose rate to the desired level. The gamma dose rate distribution for this case is shown in figure~\ref{fig5a} (c). The 3~$\upmu$Sv/h border is on the outer surface of the shielding. The neutron dose rate outside the shielding is dominated by fast neutrons from photonuclear reactions in the lead shielding (similar to the shielding for the velocity described in the last section) and is shown in figure~\ref{fig5a} (b). The neutron dose rate does not exceed 3~$\upmu$Sv/h on the outer surface of the 100\mbox{-mm}-thick lead shielding. Hence no additional shielding against neutrons is necessary in this case.

In addition to the statistical uncertainties shown in the simulated dose maps, systematic uncertainties are inherent in the Monte Carlo simulations. The main sources include evaluated nuclear data, source normalization, material-density variations, and model idealizations. However, evaluating the shielding under a conservative 60~MW scenario with a misaligned S-bender polarizer and a static velocity selector, intrinsically provides a factor-of-two power margin. This substantial margin effectively compensates for these underlying calculation uncertainties.

\section{Mechanical Design} 
\subsection{Component supports in the beam line}

\begin{figure}[!htbp]
	\centering
	\includegraphics[width=0.58\columnwidth]{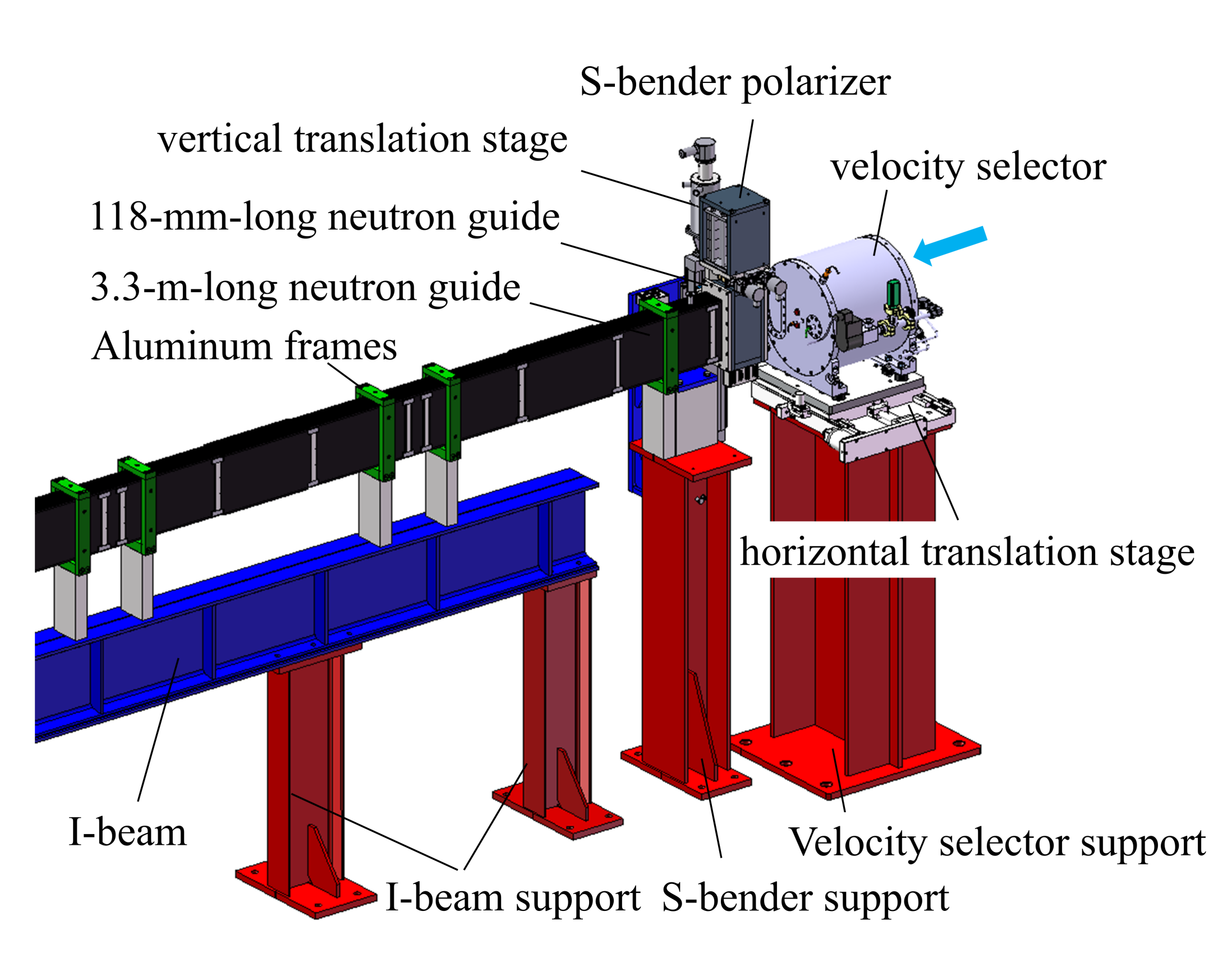}
	\caption{3D schematic layout of the beam line key components: the velocity selector, S-bender polarizer, and neutron guides. Concrete foundation below the support is hidden. The incoming neutron beam is depicted by blue arrow.}
	\label{fig8}
\end{figure}
The mechanical layout is structured to facilitate precise positioning of the beam line components shown in figure~\ref{fig8}. At the upstream end, a specialized stand is employed to support the horizontal translation stage of the velocity selector. Downstream, the design integrates the S-bender polarizer and a 118\mbox{-mm} neutron guide onto a shared vertical translation stage. For the subsequent 3.3\mbox{-m} neutron guide, a free-standing vacuum tube configuration is adopted, sealed at both ends with 0.5\mbox{-mm}-thick aluminum windows, which eliminates the need for a heavy external vacuum housing. This guide comprises three 1.1\mbox{-m}-long sections sealed with silicone and features an adjustment system utilizing precision spring screws within the aluminum frames to achieve a fine alignment accuracy of 0.01~mm at the beam center.

To meet non-magnetic requirements, 304 stainless steel is selected as the primary material for all supports and the I-beam. The interface between the supports and the floor is designed with a 30~mm clearance to allow for post-alignment grouting. This strategy is specifically intended to accommodate floor level variances of approximately 10~mm resulting from historical foundation settlement. Furthermore, the support modules incorporate a dual-mechanism for precise installation. Vertically, leveling screws threaded through the bottom plates engage with recessed base plates placed underneath the supports, allowing for fine-tuning within a range of $\pm 0.5$~mm. Horizontally, the design employs a combination of circular and slotted mounting holes for chemical anchor bolts. This slot configuration serves to compensate for installation tolerances and prevent binding, ensuring that the target positioning accuracy of $\pm 0.5$~mm along the beam axis is reliably achieved. To facilitate this precise positioning, scribing lines are integrated onto each support to serve as visual alignment targets along the beam axis.

\subsection{The velocity selector and S-bender polarizer shielding}

\begin{figure}[!htbp]
	\centering
	\includegraphics[width=0.58\columnwidth]{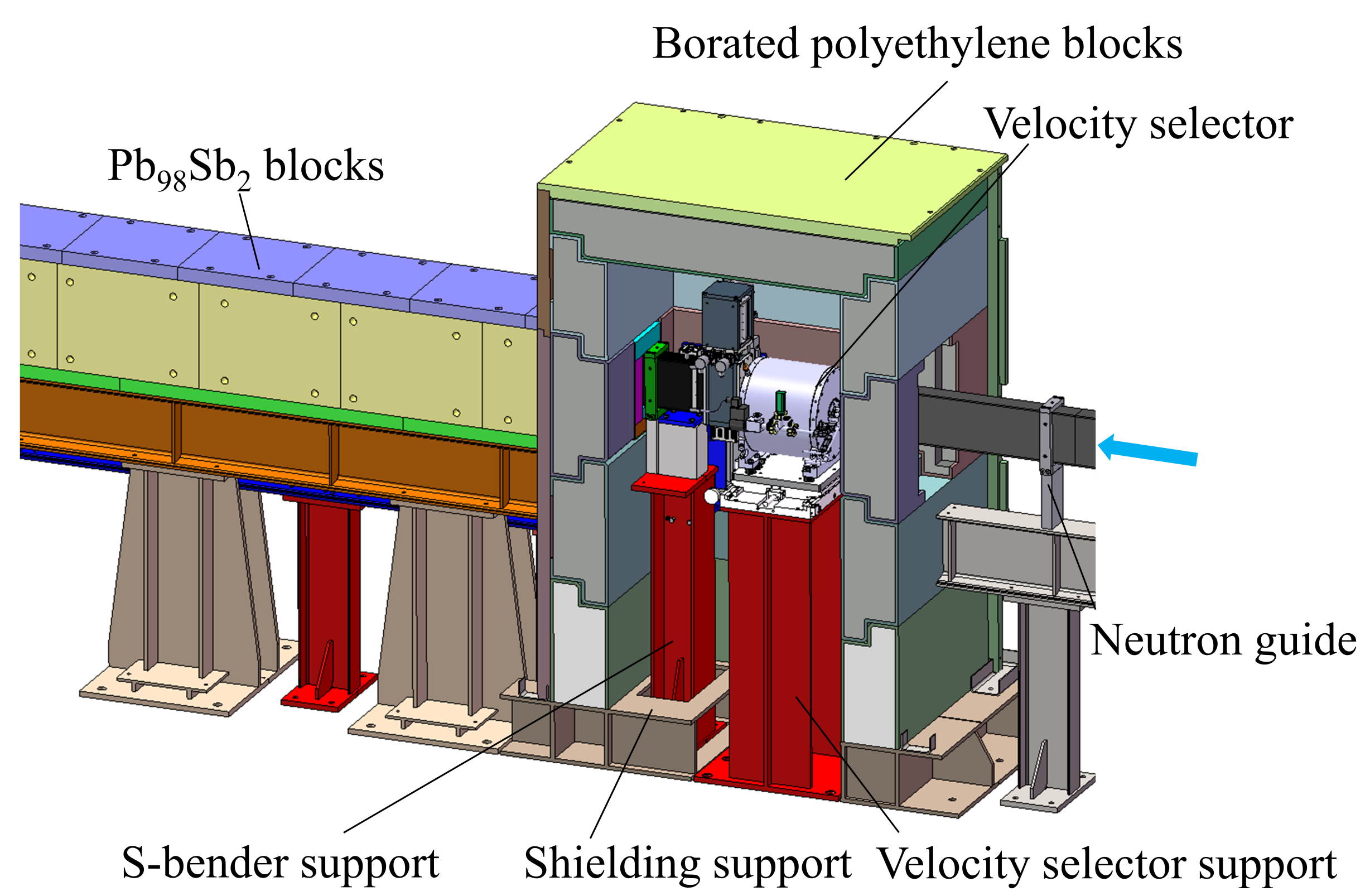}
	\caption{The construction and the main components of the velocity selector shielding. The boron carbide rubber mat inside and part of borated polyethylene pieces are hidden. Concrete foundation below the support is hidden. The incoming neutron beam is depicted by blue arrow.}
	\label{fig9}
\end{figure}
As illustrated in figure~\ref{fig9}, the shielding assembly for the velocity selector and S-bender polarizer employs a composite layered structure, consisting of a 5\mbox{-mm}-thick boron carbide rubber mat, a 224\mbox{-mm}-thick lead core, and an outer 50\mbox{-mm}-thick BPE enclosure. To facilitate fabrication, installation, and future maintenance access, the lead house is modularized into six manageable blocks. These blocks are joined using a labyrinthine stepped design, specifically, two 90$^{\circ}$ bends with a 122~mm offset, to effectively minimize radiation leakage through the seams. Structurally, each block acts as a composite unit, fabricated by casting liquid lead into a welded casing of 12\mbox{-mm}-thick 304 stainless steel, achieving a total wall thickness of 224~mm. Precise manufacturing controls are applied to the mating surfaces, with tolerances defined as $-0.5$ to $0$~mm for male joints and $0$ to $+0.5$~mm for female joints. This design constraint ensures that the inter-block gaps are strictly maintained within a range of $0$--$1$~mm. Internally, a minimum clearance of 50~mm is preserved between the shielding wall and the beam line components, with the inner surfaces lined with 5\mbox{-mm}-thick boron carbide rubber.

Surrounding the lead core is a secondary neutron shielding layer composed of 50\mbox{-mm}-thick borated polyethylene (containing 5~wt.\%~B$_2$O$_3$). This enclosure comprises nine segmented blocks which also utilize a stepped joint configuration to ensure radiation tightness. Similar to the lead core, these joints feature two 90$^{\circ}$ bends, though with a smaller offset of 25~mm. The fabrication and assembly processes are strictly controlled to ensure that gap sizes between adjacent borated polyethylene segments remain ${\leq}0.5$~mm.

The shielding support structure follows a similar installation strategy to the beam line supports, featuring a 30~mm floor clearance for post-alignment grouting and fixation via chemical anchor bolts. While the positioning tolerance for the shielding is less critical than that of the optical components, the design emphasizes foundation isolation where the concrete foundation beneath the shielding support is physically decoupled from the foundations supporting the neutron guide, S-bender polarizer, and velocity selector. This structural separation is intended to minimize mutual mechanical interference, effectively preventing stress transfer caused by settlement or load redistribution while also mitigating vibration transmission.

	\begin{figure}[!htbp]
	\centering
	\begin{subfigure}[b]{0.48\textwidth}
		\centering
		\includegraphics[width=\textwidth]{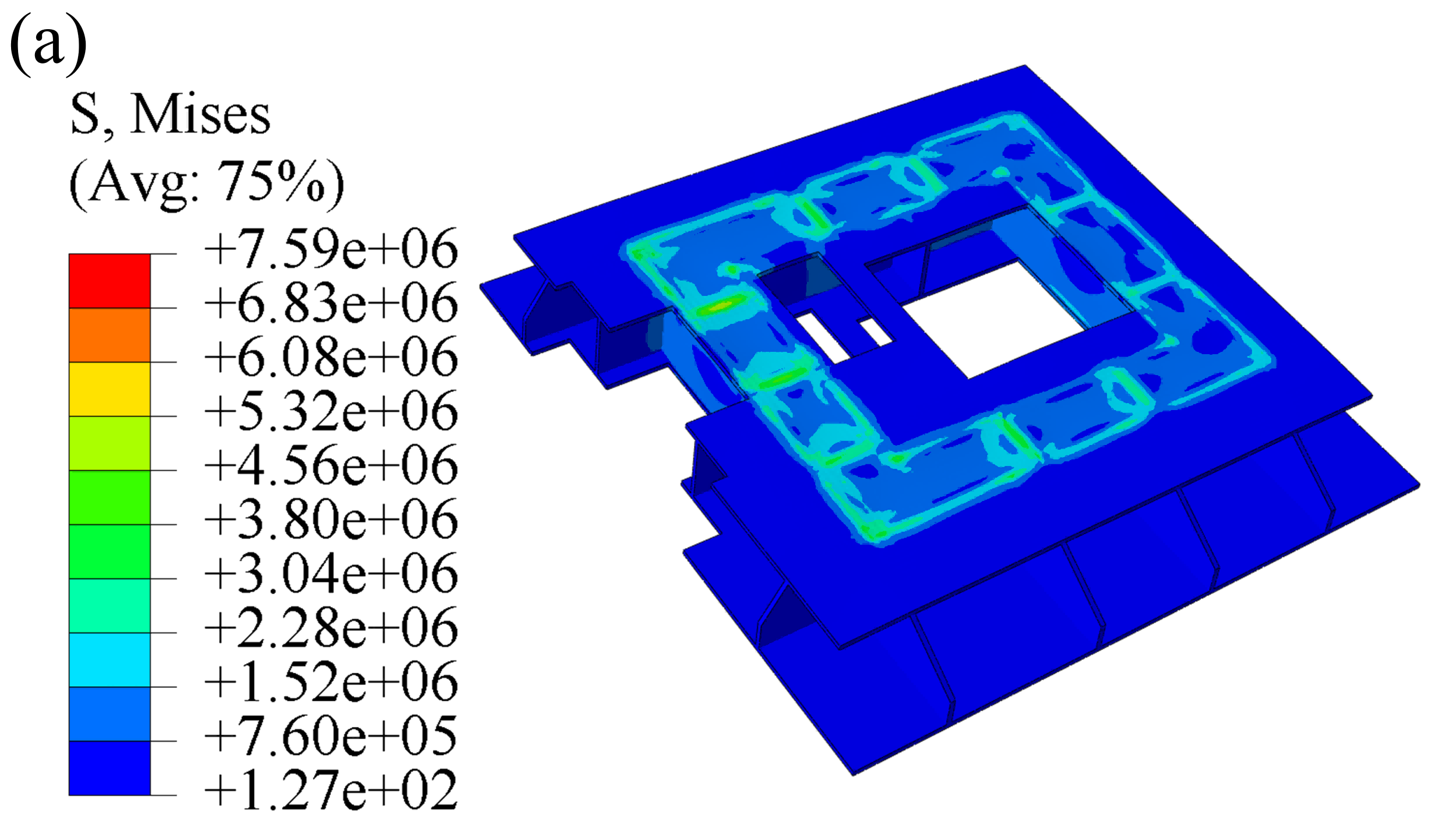}
	\end{subfigure}
	\begin{subfigure}[b]{0.48\textwidth}
		\centering
		\includegraphics[width=\textwidth]{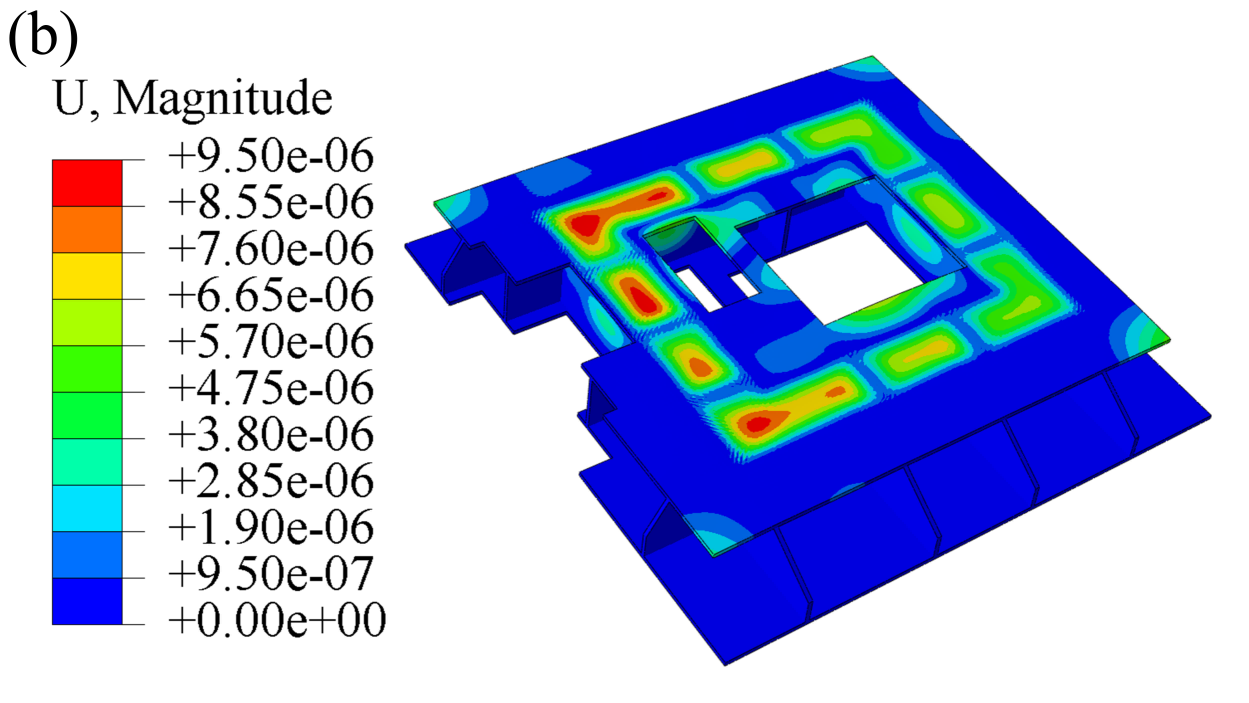}
	\end{subfigure}
	\caption{The Mises stress and deformation distributions of the velocity selector shielding support. (a) the Mises stress in the unit of Pa and (b) deformation distribution of the support in the unit of m contoured on the  support simulated from the finite element analysis. To better show the deformation, the structure is deformed by a scale factor of 1000.}
	\label{fig10}
\end{figure}
To validate the structural integrity of the support under the substantial static load of 18,200~kg, a finite element method (FEM) analysis was conducted. The support is constructed from 304 stainless steel to satisfy the non-magnetic requirement. Simulation results confirm that the maximum stress levels remain well below the ultimate yield strength of the material. Furthermore, as minimizing deformation is critical to maintaining the tight gap tolerances of the labyrinthine joints, the analysis focused on displacement. The maximum calculated deformation is 0.01~mm (see figure~\ref{fig10} (b)), confirming that the design's rigidity is sufficient to ensure shielding performance.

\subsection{The neutron guide shielding}

\begin{figure}[!htbp]
	\centering
	\includegraphics[width=0.6\columnwidth]{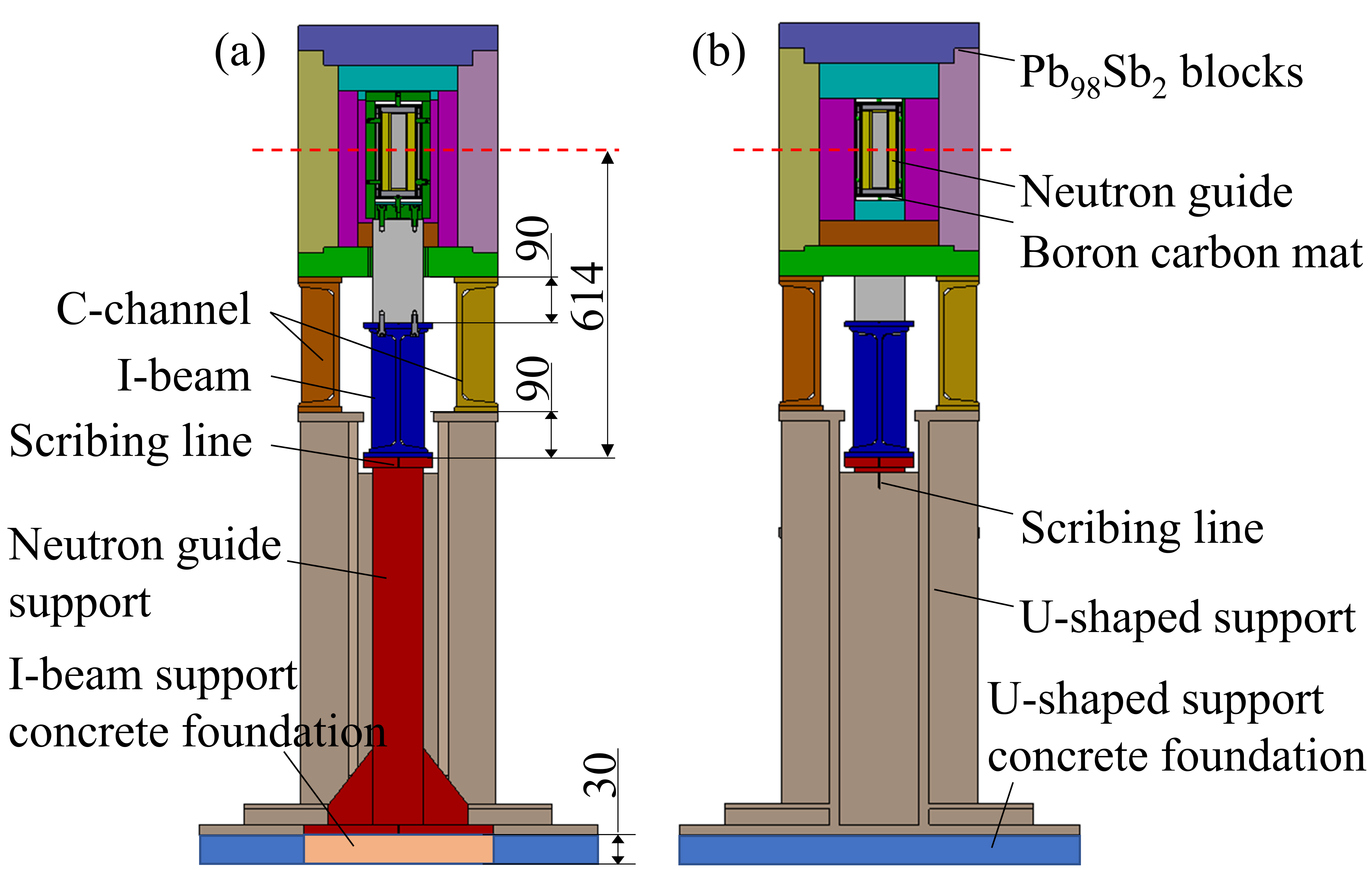}
	\caption{The construction and the main components of the neutron guide shielding. The concrete foundations for the (a) I-beam support and (b) U-shaped support are highlighted in orange and blue, respectively. The beam level is represented by the red dash lines. The key distances are marked with numbers in units of mm.}
	\label{fig11}
\end{figure}

The shielding design for the neutron guide employs a dual-layer encapsulation strategy. The primary layer consists of 5\mbox{-mm}-thick boron carbide rubber, which completely wraps the neutron guide structure. Surrounding this is a secondary shielding layer composed of 100\mbox{-mm}-thick lead-antimony alloy (Pb$_{98}$Sb$_{2}$). The inclusion of 2\% antimony (Sb) is a strategic material choice intended to increase the mechanical rigidity of the lead blocks, thereby eliminating the need for an external steel casing and simplifying the overall structure.
The realized layout of the shielding slightly differs from the simulated layout, where the boron carbide rubber mat was mounted on the inner surfaces of the lead shielding. In addition the lead shielding is not open to the floor but has a bottom below the beam axis. No additional simulations were necessary for the floor of the shielding: since neutron scattering and gamma production are symmetric around the beam axis, the radiation levels below the beam axis are the same as above the beam axis which were already simulated.

To facilitate fabrication and mounting, the Pb$_{98}$Sb$_{2}$ shielding is constructed from segmented blocks, each 500~mm in length. Consistent with the design of the velocity selector and S-bender shielding, these segments incorporate stepped labyrinth joints, featuring two 90$^{\circ}$ bends with a 25~mm offset, to prevent radiation leakage. Standard manufacturing tolerances applicable to this alloy are sufficient to achieve near-zero gap sizes between adjacent blocks. The shielding assembly is positioned with a clearance, offset 5~mm horizontally and 10~mm vertically from the inner boron carbide rubber layer, to prevent mechanical contact and also to achieve a compact structure.

The shielding assembly is mounted on a dedicated support system consisting of longitudinal C-channels seated on U-shaped supports. As shown in figure~\ref{fig11}, the U-shaped geometry is specifically designed to accommodate the central I-beam, allowing it to pass through the support structure without interference. Spatially, the upper surface of the U-shaped supports is positioned higher than the I-beam support. Here the I-beam support is located 90~mm below the U-structure. This configuration ensures a maintained gap between the Pb$_{98}$Sb$_{2}$ blocks and the I-beam. At the interface with the upstream S-bender shielding, a labyrinth-type connection is employed, requiring strict gap control within $0$--$1$~mm.

\begin{figure}[!htbp]
	\centering
	\begin{subfigure}[b]{0.48\textwidth}
		\centering
		\includegraphics[width=\textwidth]{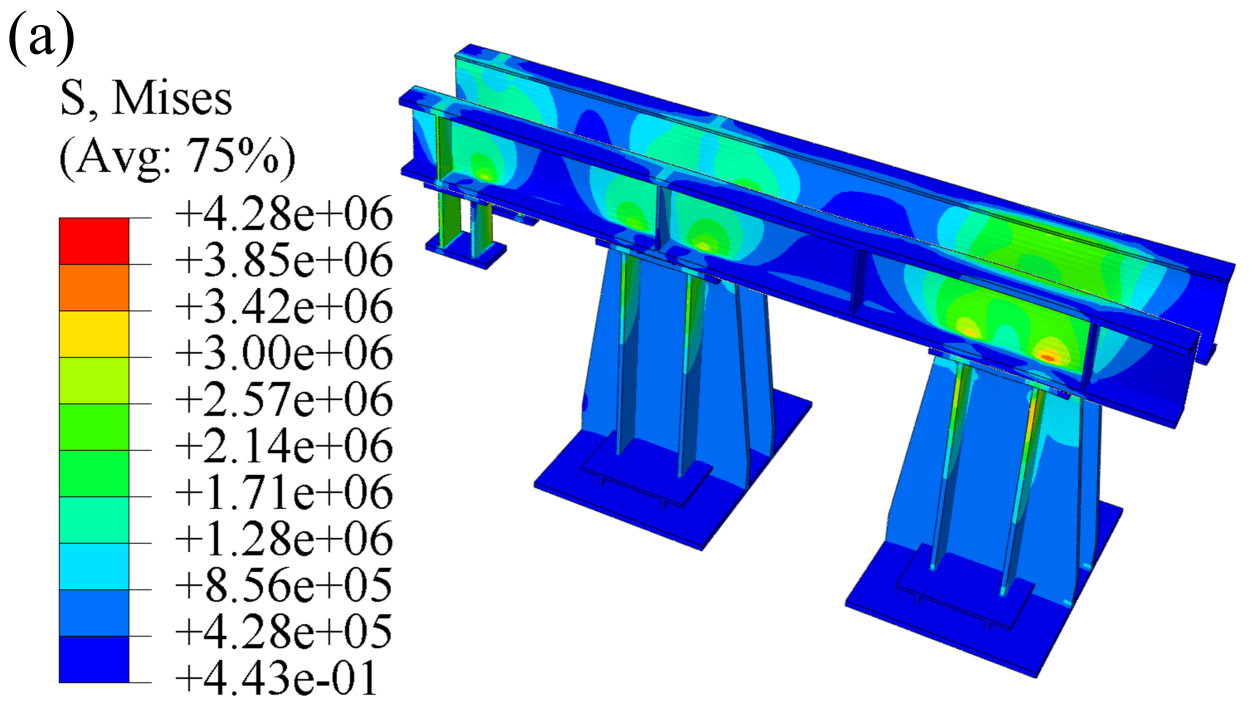}
	\end{subfigure}
	\begin{subfigure}[b]{0.48\textwidth}
		\centering
		\includegraphics[width=\textwidth]{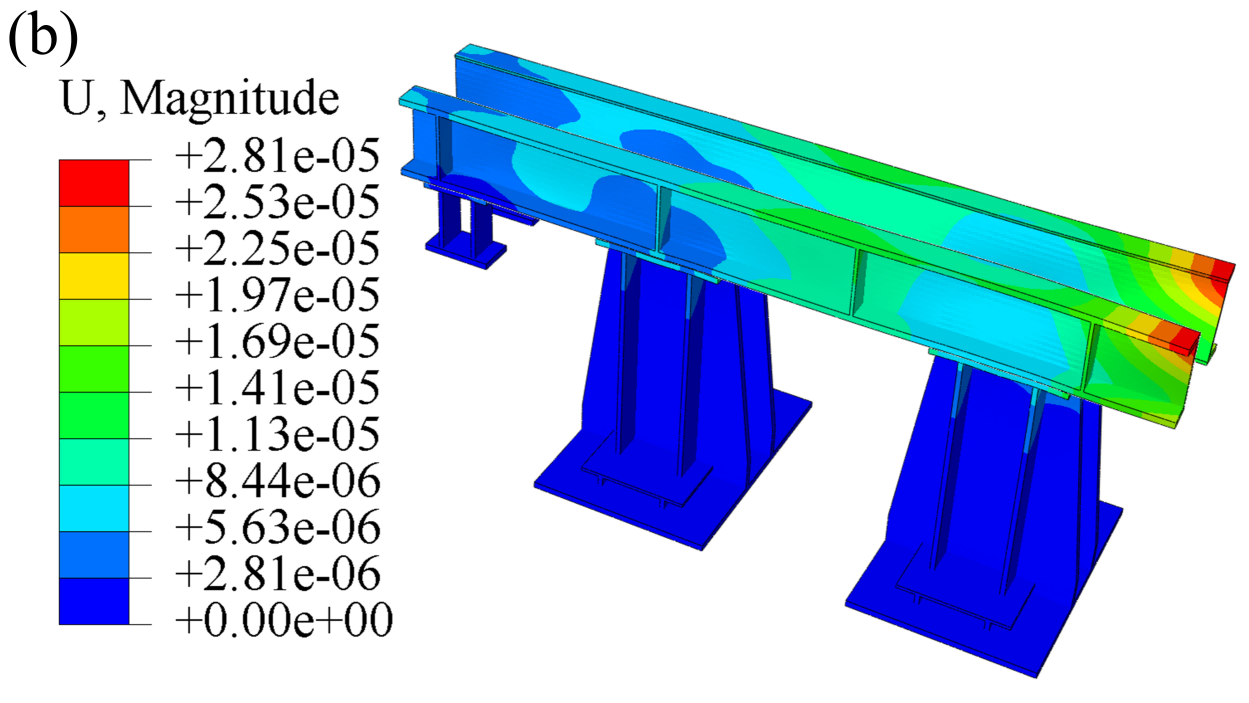}
	\end{subfigure}
	\caption{The Mises stress and deformation distributions of the neutron guide shielding support. (a) The stress in the unit of Pa, and (b) the deformation in the unit of m contoured on the C-channels and the two U-shaped supports simulated from the finite element analysis. To better show the deformation, the structure is deformed by a scale factor of 1000.}
	\label{fig12}
\end{figure}
Similar to the upstream components, the U-shaped supports, constructed from 304 stainless steel, are designed with a 30~mm floor clearance for post-alignment grouting and are secured via chemical anchor bolts. To facilitate this precise positioning, scribing lines are integrated onto U-shaped supports, see figure~\ref{fig11}. The installation targets a leveling accuracy of 0.5~mm and an axial alignment accuracy of 0.5~mm in the horizontal plane. Crucially, the concrete foundations for the shielding supports are physically decoupled from those of the I-beam supports. This foundation isolation strategy is implemented to minimize mechanical impact and stress transfer to the neutron guide system.
To validate the structural rigidity of the support under a static load of 5,200~kg, the FEM analysis was performed with fixed boundary conditions applied to the base. Simulation results confirm that stress levels remain well below the ultimate yield strength. More importantly, the maximum deformation is calculated to be 0.03~mm (see figure~\ref{fig12} (b)). This deformation is critical for maintaining the integrity of the labyrinth joints and ensuring the overall performance of the neutron guide shielding.

\section{Installation procedure}

\subsection{Installation and alignment of the support}

As the supports serve as the foundation for the neutron guide, S-bender polarizer, and velocity selector, achieving the designed installation accuracy is critical. To ensure precise positioning of the foundation interface, a template-based marking method was employed. Thin sheet metal templates, pre-drilled with specific hole patterns, were aligned on the floor to guide the installation of the chemical anchor bolts. Subsequently, the supports were roughly positioned by fitting their base plates over these pre-installed chemical anchor bolts, allowing the bolts to protrude through the mounting holes. Recessed base plates were placed beneath each support, allowing the leveling screws threaded through the support bases to engage with these plates. The vertical level of the supports was adjusted by turning these screws to meet the design accuracy requirement within $\pm 0.5$~mm, verified using a laser level.

With a design beam center height of 1400~mm above the floor, the I-beam supports were positioned to be $(614 \pm 0.5)$~mm below the beam axis. Horizontally, the supports were finely adjusted until their integrated scribing lines aligned with the beam axis to an accuracy of $\pm 0.5$~mm, as verified by a theodolite. Similarly, the U-shaped supports were positioned $(524 \pm 0.5)$~mm below the beam axis, maintaining the same horizontal accuracy of $\pm 0.5$~mm. Upon completion of the fine alignment, all anchor nuts were tightened to secure the supports, and the gap between the supports and the floor was grouted with non-shrink cement to ensure rigid fixation.

\subsection{Assembly of shielding and components}
\begin{figure}[!htbp]
	\centering
	\includegraphics[width=0.9\columnwidth]{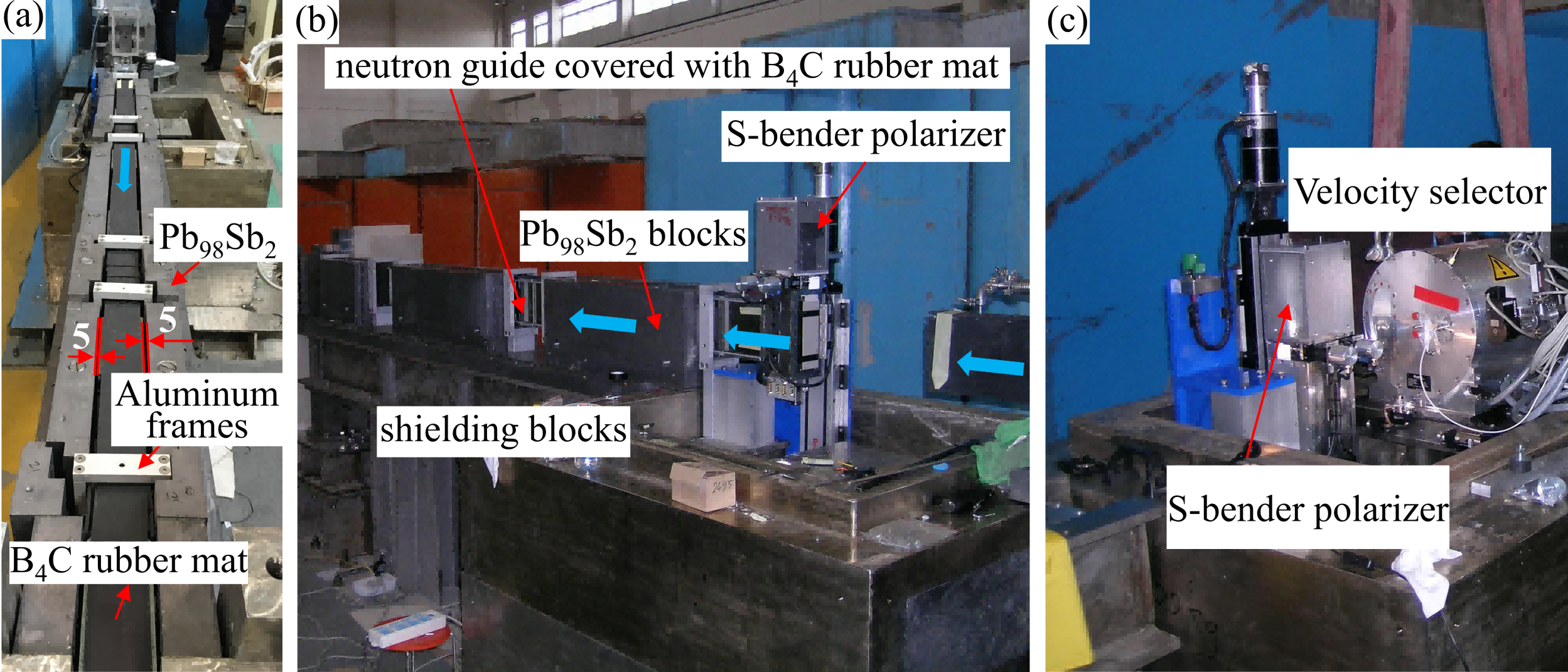}
	\caption{Shielding assembly for the neutron guide, S-bender polarizer, and velocity selector during construction. (a) Pb$_{98} $Sb$_{2}$ blocks are mounted around the neutron guide. The clearance between the neutron guide with the 5\mbox{-mm}-thick rubber mat and the lateral Pb$_{98}$Sb$_{2}$ blocks is 5~mm, consistent with the design requirements. (b) The lead shielding for the velocity selector and S-bender is partially mounted, and note that the velocity selector has been removed in this view. (c) The velocity selector and the S-bender were installed. The incoming neutron beam is depicted by blue arrow. Dimensions in mm.}
	\label{fig13}
\end{figure}

Following the fixation of the supports, the C-channels and I-beam were mounted on their respective bases, with their levels re-confirmed by a laser level. The I-beam was then adjusted horizontally to align with the beam axis within an accuracy of $\pm 0.5$~mm. Once the support infrastructure, including the I-beam and C-channels, was secured, the bottom layer of Pb$_{98}$Sb$_{2}$ blocks and the aluminum frames equipped with precision spring screws were installed first. Subsequently, the lateral Pb$_{98}$Sb$_{2}$ blocks were mounted, excluding only those surrounding the aluminum frames.

The neutron guide was then installed and precisely aligned with the beam axis to a fine accuracy of $\pm 0.01$~mm. This high precision was achieved using a theodolite and laser tracker to guide the adjustment of four pairs of spring screws located on the aluminum frame, which actuate protective plates on the guide surfaces. Finally, the remaining lateral and upper Pb$_{98}$Sb$_{2}$ blocks were installed to complete the shielding assembly. Figure~\ref{fig13} illustrates the installed shielding at the end of the CNGA beam line. From figure~\ref{fig13} (a), it is also observed that the consistent horizontal gap between the neutron guide with the 5\mbox{-mm}-thick rubber mat and the Pb$_{98}$Sb$_{2}$ blocks validates the quality of the support alignment. This successful implementation demonstrates that the compact design strategy, featuring a minimal 5~mm offset, was practically feasible, achieving a tightly integrated structure without mechanical interference.

\subsection{Radiation monitoring}
To provide an experimental check of the shielding calculation, radiation dose rates were measured during the routine 30~MW operation of the CARR facility. These measurements listed in Table~\ref{tab:validation} were conducted during the early commissioning phase, when the S-bender polarizer was translated vertically out of the beam to prioritize primary beam calibrations. The measurements therefore correspond to the polarizer-out-of-beam commissioning configuration.
The early commissioning survey did not include dedicated polarizer-in-beam benchmark measurements. During later polarized-neutron experiments, routine area dose-rate checks were performed for radiation-safety purposes. These checks were operational surveys rather than dedicated quantitative benchmarking measurements, and therefore they are not tabulated as point-wise validation data for comparison with the MCNP model.

Neutron dose-rate measurements were performed using a Berthold UMo~LB~123 universal monitor equipped with an LB~6411 spherical moderated neutron probe, calibrated for neutron ambient dose equivalent rate, $\dot{H}^*(10)$. Gamma dose-rate measurements were performed using a Thermo Scientific RadEye~G-10 Geiger--M{\"u}ller survey meter, calibrated for gamma ambient dose equivalent rate, $\dot{H}^*(10)$.
The uncertainties shown with the measured values in Table~\ref{tab:validation} represent the standard deviations of repeated on-site readings. A conservative detector-related systematic uncertainty of $\pm 30\%$ was assigned to the survey readings to account for calibration and energy-response effects. The MCNP relative statistical uncertainties are listed
separately in Table~\ref{tab:validation}.

\begin{table}[htbp]
	\centering
	\small
	\setlength{\tabcolsep}{2pt} 
	\renewcommand{\arraystretch}{1.2} 
	
	\caption{Comparison of measured ambient dose equivalent rates during the
		early 30~MW commissioning survey with the S-bender polarizer translated out
		of the beam and corresponding MCNP values normalized to 30~MW.}
	\label{tab:validation}
	
	\begin{tabularx}{\textwidth}{l cc *{4}{>{\centering\arraybackslash}X}}
		\hline
		\textbf{Position} &
		\textbf{Radiation} &
		\textbf{Detector} &
		\textbf{Measured $\dot{H}^*(10)$ @30~MW ($\mu$Sv/h)} &
		\textbf{Detector syst. uncertainty} &
		\textbf{MCNP $\dot{H}^*(10)$ @30~MW ($\mu$Sv/h)} &
		\textbf{MCNP rel. stat. uncertainty} \\
		\hline
		A & Gamma  & RadEye~G-10    & $0.50 \pm 0.16$ & $\pm 30\%$ & 0.51 & $4.6\%$ \\
		B & Gamma  & RadEye~G-10    & $0.50 \pm 0.05$ & $\pm 30\%$ & 0.52 & $4.4\%$ \\
		C & Neutron & LB~123+LB~6411 & $0.17 \pm 0.01$ & $\pm 30\%$ & 0.20 & $8.7\%$ \\
		\hline
	\end{tabularx}
	
	\vspace{1ex}
	\begin{minipage}{0.98\linewidth}
		\footnotesize
		\setlength{\parindent}{0pt}
		\textit{Note:} The uncertainties shown with the measured values represent the standard deviations of repeated on-site readings. A conservative detector-related systematic uncertainty of $\pm 30\%$ is listed separately.
	\end{minipage}
\end{table}

The three measurement positions were selected as representative accessible survey locations around the velocity-selector shielding, according to the calculated polarizer-out-of-beam dose-rate distribution and practical survey conditions.

As shown in Table~\ref{tab:validation}, the measured and calculated dose-rate magnitudes at the selected accessible positions are consistent within the measurement and MCNP uncertainties. All measured values are below the internal dose-rate guideline. The polarizer-in-beam shielding sufficiency is assessed using the MCNP calculations. Together, the polarizer-out-of-beam commissioning benchmark, the factor-of-two power margin relative to routine 30~MW operation, the uncertainty discussion above, and the operational access controls described in Section~2, provide a conservative basis for evaluating shielding sufficiency under routine operation. The practical reliability of the shielding design has also been supported by six years of routine operation. The neutron shielding design has been adopted by the HETU instrument at CMRR~\cite{Wang:te5115}.

\section{Conclusion}

The design and optimization of the shielding for the velocity selector, S-bender polarizer and neutron guide have been successfully accomplished. Through comprehensive Monte Carlo simulations and purpose-built mechanical design, effective shielding solutions with optimized tolerances were developed using a compact sandwich structure of B$_{4}$C rubber, lead, and borated polyethylene. This design effectively addresses radiological challenges, while complying with strict space limitations and floor load constraints.
The Pb$_{98}$Sb$_{2}$ alloy provides the necessary rigidity for self-support. It satisfies the tolerance requirements for the gaps between mating blocks. An installation and alignment method has been developed to ensure the co-alignment of the various spectrometer and shielding components.
This work represents a significant advancement in neutron instrumentation in China and offers valuable insights and methodologies that can serve as a reference for future developments in the field.

\section{Acknowledgements}

Many people contributed to the design and construction of the shielding. The authors are particularly indebted to Junfeng Li and Wenwei Shen. 
The authors also gratefully acknowledge the help from Dr. P. Link from Heinz Maier-Leibnitz Zentrum, Technische Universität München, and Dr. A. Schneidewind from the Jülich Center for Neutron Science (JCNS), Forschungs zentrum Jülich GmbH. 
This work is supported by NSAF Joint Fund of National Natural Science Foundation of China (No. U2030106), and the Special Fund for Research on National Major Research Instruments of NSFC (No.11227906), and the National Natural Science Foundation of China (No. 12304185).


\end{document}